# Modeling changes in atomic structure around a vacancy with increasing temperature and calculation of temperature dependences of vacancy characteristics in bcc iron

*M. Boboqambarova[1*], A. V. Nazarov[1,2]*
[1] National Research Nuclear University MEPhI, Moscow, Russia
[2] National Research Center "Kurchatov Institute", Moscow, Russia
[*] e-mail: boboqambarovam@gmail.com

We developed an original natural thermostat algorithm to simulate the direct change in interatomic distances with temperature in both an ideal crystal and a system with a vacancy. In contrast to previous work, the results indicate that in a system with a defect, the radii of the first ten coordination spheres change nearly linearly with increasing temperature. However, the coefficients determining these dependencies, unlike the interatomic distances farther from the vacancy, are not equal to the linear thermal expansion coefficient of the perfect crystal. Knowing the atomic coordinates and their temperature evolution, we calculated the temperature dependence of the vacancy formation energy and the vacancy relaxation volume in bcc iron. The significance of these effects for the accurate calculation of atomic diffusion coefficients is analyzed and discussed.

## I. INTRODUCTION

Point defects govern microstructure evolution and degradation in metals, fundamentally influencing their properties [1]. Therefore, to predict kinetics, detailed knowledge of the characteristics of such defects and the possible dependence of these characteristics on temperature is needed. In this context, many studies have focused on defects, the results of which are presented in numerous monographs [1-8] and key publications [9-21].

The equation for the equilibrium concentration of vacancies is often written as follows [1]:

$$c_V = A exp\left(-\frac{E^f}{kT}\right) \quad (1)$$

where $E^f$ is the vacancy formation energy and $A$ is the pre-exponential factor related to the vacancy formation entropy $S_V^f$ [1]:

$$A = exp\left(\frac{S_V^f}{kT}\right) \quad (2)$$

Typically, when analyzing experimental data (or results from molecular dynamics-MD simulations), the Arrhenius plot is used to determine the vacancy formation energy. In many cases, the dependence of $Ln(c_V)$ on the inverse temperature is linear. However, as emphasized in the monograph [2] and in the article [9] by Girifalco, the linearity of the $Ln(c_V)$ dependence on the Arrhenius plot does not guarantee that the energy $E^f$ is independent of temperature. If this energy depends linearly on temperature, the dependence on the Arrhenius plot will still appear as a straight line.

As shown in the work by Glyde [8], owing to the increase in interatomic distances with increasing crystal temperature, the defect characteristics calculated at the atomic level using interatomic potentials should change. This is the first problem that needs to be addressed for more accurate calculations of defect characteristics in metals and alloys, which is necessary for fully implementing the multiscale modeling concept. The second issue is reducing the negative effects associated with the periodic boundary conditions used in many algorithms for modeling point defects and atomic jumps. This is discussed in detail, for example, in the works by Ma and Dudarev [19,20], where one solution was proposed. However, it should be noted that in earlier works by Johnson [10], a model was developed that does not use periodic boundary conditions but accounts for defects being far apart with minimal mutual influence on the structure. In addition to Johnson's works and those of his coauthors, the approach he proposed is used by other authors [13].

The algorithms used in these works avoid problems caused by periodic boundary conditions but fail to achieve satisfactory convergence of iterations for large numbers of atoms in the simulation cell. Additionally, molecular statics (MS) are implemented at zero temperature, so possible temperature dependencies of characteristics are not accounted for. In our previous works [22-24], a modified molecular statics (MMS) algorithm was proposed, which retains the advantages of Johnson's model while ensuring good convergence of the iterative procedure for large systems, particularly in modeling the atomic structure near pores [25-27], where the number of particles reached $10^6$.

The temperature dependence of the defect characteristics in aluminum has been considered in works [11-12]; however, effects related to periodic boundary conditions have not been addressed. Similar modeling of defects in iron was conducted [18], with a main focus on the magnetic contribution to the diffusion activation free energy. The thermal expansion contribution in Ni is taken into account through the lattice parameter variation in [14]. The temperature dependence of the vacancy formation free energy was calculated via MD simulations [16, 21], where periodic boundary conditions were also used. Moreover, it was noted in Ref. [16]: "this calculation neglects the fact that lattice regions in the vicinity of the defect may have local thermal expansion factors somewhat different from the expansion factor of the perfect lattice". In our previous works [28-31], we studied changes in the atomic structure near defects via the MD cluster model (Model with the free-boundary conditions). Preliminary results suggested that with increasing temperature, the distance from the defect center to the coordination spheres changes proportionally to the temperature dependence of the lattice parameter. However, new calculations have shown that such proportionality holds with high accuracy only for atoms far from the defect.

In this work, thermal expansion is modeled for bcc iron by observing the direct change in interatomic distances with temperature in both an perfect crystal and a system with a vacancy. We use the natural thermostat (NT) [32-34], which is based on a combination of MMS [22-24] and MD, and is different from the Nosé–Hoover thermostat, and the Berendsen thermostat. NT allows the avoidance of problems with periodic boundary conditions and more accurately accounts for changes in interatomic distances with increasing temperature than our previous methods did [28-31]. Additionally, the natural thermostat makes it possible to account for, unlike in [14], the different physical nature of thermal expansion and the mechanically induced increase of the lattice constant due to stretching. If the atomic structure is known at a given temperature, the defect characteristics can be calculated and, consequently, their temperature dependence can be determined. As a result, we found the aforementioned dependencies for the vacancy formation energy and relaxation volume in bcc iron.

## II. METHODS

### A. MD. The free-boundary conditions

In the first part of the work, we studied the thermal expansion of an ideal crystal. The MD cluster model was used, as in [28-31], with the EAM potential by Ackland [35]. The coordinates of the central atom and its neighbors inside the nearest shells are gathered during a simulation run and then averaged. Via mean positions, we obtain interatomic distances, and then via the geometry of the BCC structures, we directly obtain the lattice parameter. Thus, the lattice parameter is determined in a similar way as it is in X-ray measurements.

The model was validated by testing for computational size and simulation time dependence. We found that the results converge to within a constant value for simulation times exceeding 1000 atomic vibrations and for computational cells containing approximately 30,000 atoms.

The dependence of the lattice constant on temperature is shown in Fig. 1.

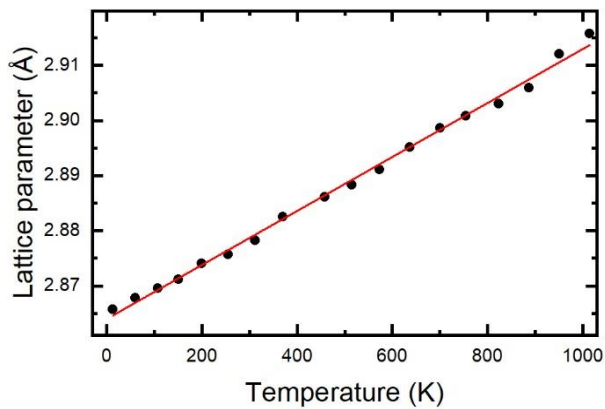

**Figure 1.** Lattice constant of bcc Fe as a function of temperature obtained via MD simulations.

The lattice constant linearly expands with increasing temperature. The calculated thermal expansion coefficient ($\alpha = 1.67 \cdot 10^{-5} K^{-1}$) agrees well with the experimental values [36], and similar results ($1.64 \cdot 10^{-5} K^{-1}$) were previously obtained [28-30].

## B. Natural Thermostat model

To study the structure around the vacancy and its changes with increasing temperature, NT was applied, which was previously developed for studying the diffusion jumps of carbon atoms in bcc iron [32, 34] and was recently used to investigate the effect of pressure on vacancy migration [34].

In this model, the simulation cell is spherical and consists of 3 regions (Fig. 2). The trajectories of the atoms in the first and second regions are determined via MD, whereas those of the atoms in the third region are fixed.

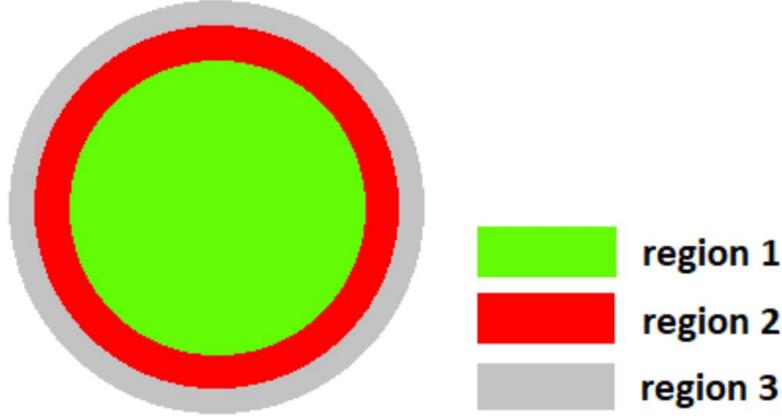

**Figure 2.** Schematic representation of the natural thermostat.

Region I is the main simulation cell, from whose averaged atomic coordinates, the vacancy characteristics are calculated; Region II is the thermostat, a thermal reservoir used to maintain a specific temperature in the simulation cell; and Region III is the elastic medium in which the atoms are embedded. Their displacements relative to perfect lattice sites are determined via solutions from isotropic elasticity theory for a spherically symmetric defect:

$$\vec{u} = C \frac{\vec{r}}{r^3}, \quad (3)$$

where $\vec{u}$ is the displacement vector, $\vec{r}$ is the vector from the vacant site to the atom in the third region, and $C$ depends on temperature [30]:

$$C(T) = C_0 \cdot (1 + \alpha T)^3, \quad (4)$$

where $C_0$ is obtained preliminarily by MMS and where $\alpha$ is found in the first stage.

To integrate the equations of motion, the velocity Verlet algorithm is used:

$$\boldsymbol{r}(t + \Delta t) = \boldsymbol{r}(t) + \boldsymbol{v}(t)\Delta t + \frac{1}{2}\frac{\boldsymbol{F}(t)}{m}\Delta t^2,$$

$$\boldsymbol{v}(t + \Delta t) = \boldsymbol{v}(t) + \frac{1}{2}\left(\frac{\boldsymbol{F}(t) + \boldsymbol{F}(t + \Delta t)}{m}\right) \cdot \Delta t, \quad (5)$$

where standard MD notations are used.

The atoms in region I directly interact with the thermostat. To maintain a specific temperature in the system, at each time step, the velocities of each atom in region II are corrected by a factor $\beta$ (as in the isokinetic variant of MD):

$$\beta = \frac{\frac{3}{2}N_2 kT}{\sum_i \frac{m_i v_i^2}{2}}, \quad (6)$$

where $i$ are indices of atoms in region II, and $N_2$ is the number of atoms in region II.

The initial velocities of the atoms in the first and second regions are set according to the Maxwell distribution for the experimental temperature.

Testing of the model showed [31-32] that in the first region, thermal equilibrium is naturally established over time, corresponding to some atomic vibrations, and that the atomic velocities satisfy the Maxwellian distribution. Since the results should not depend on the simulation cell size, testing revealed that the condition is met with sufficient accuracy at the following parameters:

$r_I$= 12a, $n_I$=14361, $r_{II}$=15a, $n_{II}$=14094, $r_{III}$=18a, $n_{III}$=19394.

During simulation, after reaching equilibrium in the system, atom coordinates are gathered during a simulation run and then averaged. Then, the average values of the radii for the first ten coordination spheres ($R_1$, $R_2$, $R_3$, $R_4$, ) are found. Similar experiments were conducted for each temperature in the range of interest.

### III. RESULTS AND DISCUSSION

The temperature dependences of the radii of the first four coordination spheres are presented in Fig. 3.

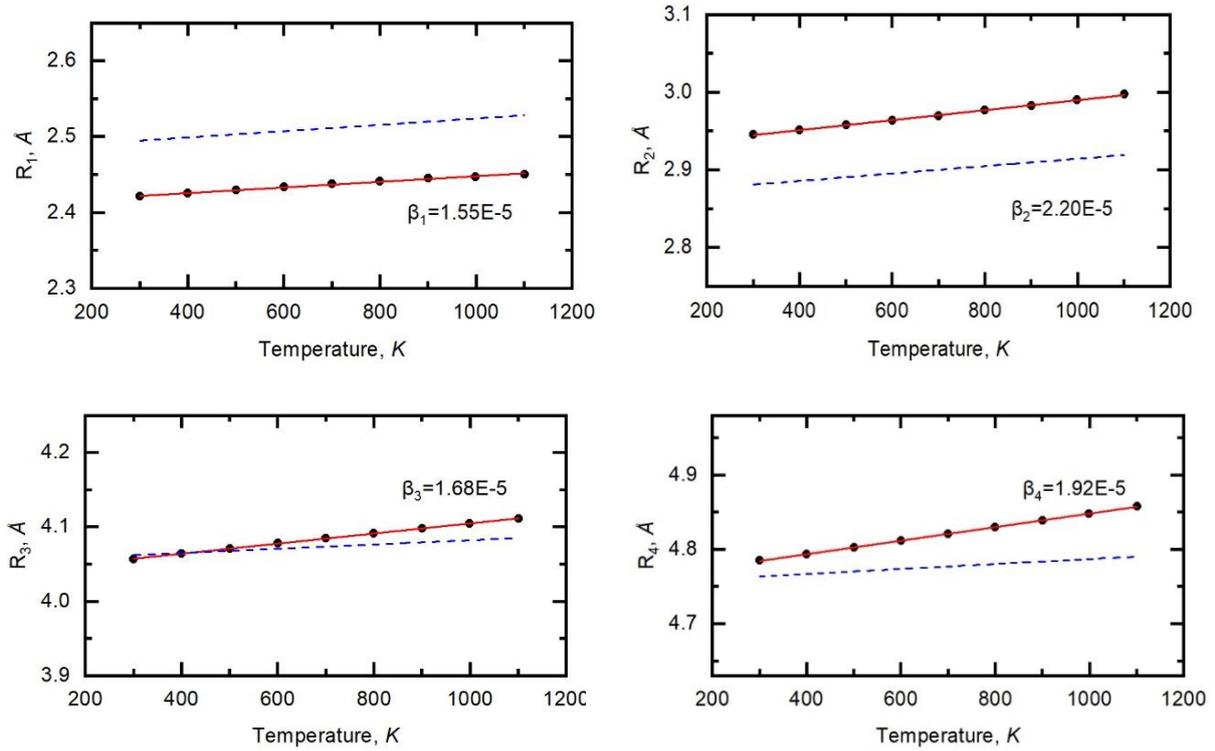

**Figure 3**. Temperature dependence of the radii of the first four coordination spheres. The dashed lines represent the corresponding temperature dependence for an ideal crystal.

Similar data were obtained for the other spheres. The next figure shows ratios of coefficients $\beta_i$ for corresponding coordination sphere radii in the system with a vacancy and in the perfect system.

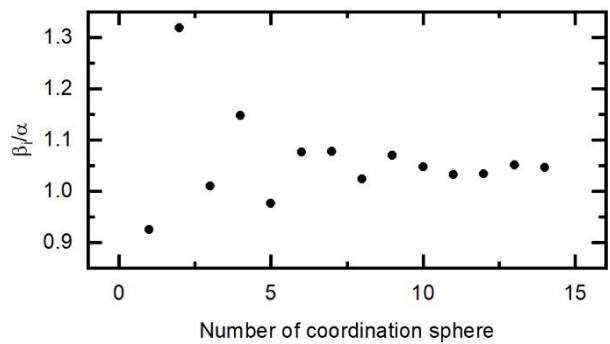

**Figure 4.** Ratios of coefficients $\beta_i$ for corresponding coordination sphere radii in the system with a vacancy and in the perfect system.

The results reveal that the radii of the first ten coordination spheres expand approximately linearly with temperature. However, the expansion coefficients β$_i$ for these spheres differ from the bulk thermal expansion coefficient, asymptotically approaching it with increasing coordination sphere number.

The value to which the results converge on the graph $\alpha = 1.69 \cdot 10^{-5} K^{-1}$. Its slight difference from the coefficient used for calculating atomic displacements in region III is likely due to the use of different ensembles: NVE in the cluster model and NPT in NT. Furthermore, the influence of the Laplace pressure is excluded in NT.

Note that distances to atoms beyond the tenth coordination sphere change with increasing temperature in proportion to the change in the lattice parameter. That is, the geometric similarity in the arrangement of such atoms is preserved.

This result is important because it allows simplification of the calculation of the average coordinates of atoms beyond the first ten coordination spheres at any temperature, considering that the constant $C$ in this case, as previously shown in our works [29-30], is found via a simple formula:

$$C(T) = u(R,T) \cdot R^2(T) = u(R, 0) \cdot (1 + \alpha T) \cdot R^2(0) \cdot (1 + \alpha T)^2 = C_0 \cdot (1 + \alpha T)^3 \quad (7)$$

Thus, knowing the coefficients $\beta_i$ for the first ten coordination spheres and using equation (7) to calculate the displacements of the remaining atoms in the system, one can recalculate the coordinates of all the atoms for any temperature based on preliminary MMS data.

When calculating the vacancy formation energy at a certain temperature, we used standard equations accepted in MS and recalculated the coordinates of the atoms corresponding to the ground state of the system for that temperature, as interpreted in Glyde's work [8], based on statistical mechanics methods. The results are shown in Fig. 5, and for comparison, data from modeling in works [29, 30] are presented, where it was assumed that interatomic distances change proportionally to the lattice parameter with increasing temperature.

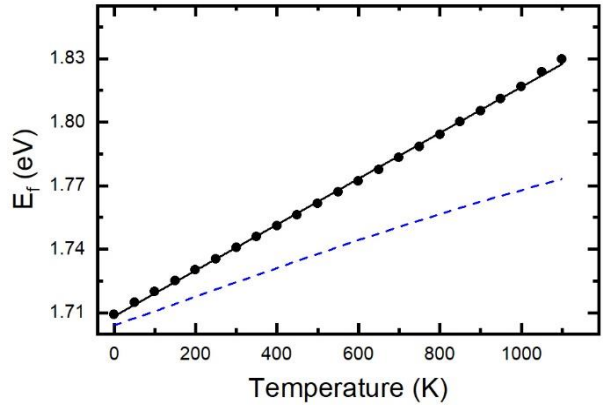

**Figure 5**. Vacancy formation energy as a function of temperature.

Figure 5 shows that the vacancy formation energy in bcc iron increases approximately linearly with temperature.

The vacancy concentration is determined by Eq. (1). According to the simulating data, the vacancy formation energy linearly depends on the temperature:

$$E^f = E_0^f + \gamma T, \quad (8)$$

where $E_0^f = 1.71 eV$ is found via MMS modeling. After substituting Eq. (8) into Eq. (1),

$$c_V = A exp(-\frac{\gamma}{k}) exp(-\frac{E_0^f}{kT}) \quad (9)$$

A similar expression is given in Glyde's article [8]. In agreement with the simulation results, the additional factor contributing to the pre-exponential is as follows:

$$f_\gamma = exp\left(-\frac{\gamma}{k}\right) = 0.285, \quad (10)$$

and differs significantly from the result reported in [28]

$$f_\gamma = exp\left(-\frac{\gamma}{k}\right) = 0.479. \tag{11}$$

Typically, when analyzing experimental data (or MD simulation results), an Arrhenius plot is used to find the formation energy. From equation (9), it follows that in such a procedure, $E_0^f$ is found, and the pre-exponential factor additionally contains a contribution defined by Eq. (10), which significantly reduces it compared with results from MS modeling, that do not account for the temperature dependence of the vacancy formation energy. The atomic diffusion coefficient is commonly evaluated using the standard equation [5]:

$$D_A(T) = c_{eq} D_V(T) f, \tag{12}$$

where $f$ is the correlation factor.

If Eqs. (9) and (10) are taken into account, we obtain

$$D_A(T) = A f_\gamma exp\left(-\frac{E_0^f}{kT}\right) D_V(T) f, \tag{13}$$

It should be emphasized that for BCC structures the correlation factor $f = 0.727$, the effect due to the temperature dependence of the formation energy, as reflected by the factor $f_\gamma$, has a significantly greater impact on reducing the self-diffusion coefficient than correlation effects.

Since the relaxation volume $V_{Rel}$ is proportional to the constant $C$ [10],
$$V_{Rel} = 4\pi C, \tag{14}$$
From Eq. (4), we obtain
$$V_{Rel}(T) = V_{Rel}(0) * (1 + \alpha T)^3 \tag{15}$$
Figure 6 shows the dependence of this characteristic on temperature.

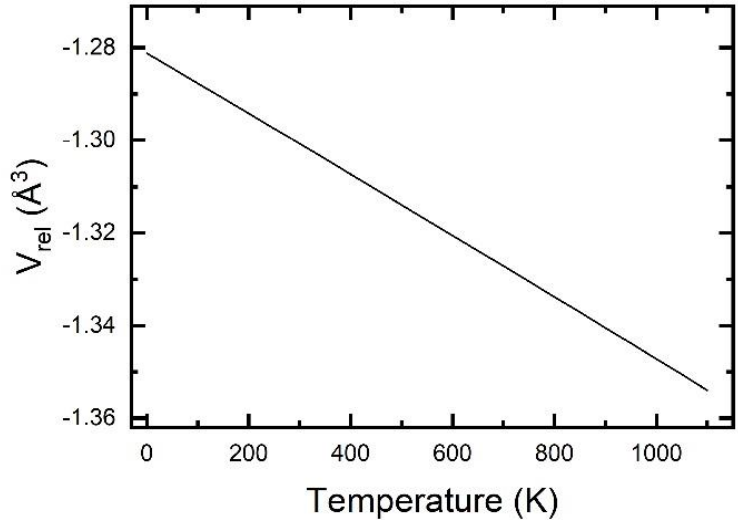

**Figure 6.** Temperature dependence of the relaxation volume.

$V_{Rel}(0)$ was previously obtained by the MMS
$$V_{Rel}(0) = -1.281 \text{ Å}^3$$

Because the atomic volume $\Omega$ is proportional to $a^3$, where $a$ is the lattice constant, the ratio $V_{Rel}(T)/\Omega(T)$, which is frequently used in the analysis of experimental data or in the modeling of the obtained quantities, does not depend on temperature.

It should also be noted that the void growth rate equation derived from the Lifshitz–Slezov theory [37] for low void density reads

$$\frac{dR}{dt} = c_{eq} D_V \frac{1}{R}\left(\Delta + 1 - exp\left(\frac{2\gamma V^f}{kTR}\right)\right), \quad (16)$$

where $\Delta = \frac{c_m - c_{eq}}{c_{eq}}$ is the vacancy supersaturation, $c_{eq}$ is the equilibrium vacancy concentration, $\gamma$ is the surface energy, and $V^f$ is the vacancy formation volume. The right-hand side of this equation includes parameters calculated in this work and in Ref. [34] by modeling techniques. These parameters also appear in equations accounting for the influence of elastic fields induced by void on the vacancy flux and the void growth rate [27]. We emphasize that void growth is a key factor in the swelling of nuclear reactor materials, and knowledge of the values of the defect characteristics and their temperature dependencies is essential for predicting these detrimental processes.

## IV. CONCLUSIONS

In this work we employed an original model (natural thermostat) that enables the investigation of temperature-dependent changes in interatomic distances in both perfect bcc crystals and systems containing vacancies. We found that changes in interatomic distances near a vacancy differ significantly from those in the perfect lattice, and we calculated the coefficients $\beta_i$, that characterize these changes. Using the atomic structure of bcc iron at a given temperature, we calculated the vacancy formation energy. While the temperature dependence of the vacancy formation energy is often neglected, our simulations, as well as previous studies [16], indicate that this effect is rather significant. However, unlike Refs [16] and [31], we take into account that the coefficients $\beta_i$ differ from the thermal expansion coefficient of the defect-free crystal. This difference arises from the fact that the atomic arrangement in the vicinity of a vacancy deviates from that in the perfect lattice, and therefore, the coefficients determined the anharmonicity of atomic vibrations differ between these structures.

The results show that the vacancy formation energy varies nearly linearly with increasing temperature, contributing an additional term to the pre-exponential factor in the equation for the equilibrium vacancy concentration. Accounting for this effect alters the estimated vacancy concentration several times. Since the atomic diffusion coefficient is proportional to the vacancy concentration, its value is therefore several times lower than that obtained from simulations that neglect the temperature dependence of the vacancy formation energy. This effect ($f_\gamma = 0.285$) is more significant than correlation effects ($f = 0.727$ for bcc).

Furthermore, in contrast to previous studies, our approach allowed to determine the temperature dependence of the vacancy relaxation volume.